\documentclass[twocolumn,aps,runinaddress,superscriptaddress,floatfix,showpacs]{revtex4}
\usepackage{epsfig}
\usepackage{rotating}
\usepackage{psfrag}
\usepackage{eepic}
\usepackage[latin1]{inputenc}
\usepackage[T1]{fontenc}
\usepackage{algorithm}
\usepackage{algorithmic}
\usepackage{color}
\newcommand{\MVM}{MSG-VM }
\newcommand{\rc}[1]{{\color{black} #1}}

\floatstyle{plain}
\newfloat{mybox}{thp}{lp}
\floatname{mybox}{Algorithm}

\begin{document}
\title{Efficient modularity optimization by\\ multistep greedy algorithm and vertex mover refinement}
\author{Philipp Schuetz and Amedeo Caflisch}
\affiliation{Department of Biochemistry, University of Zurich, Winterthurerstrasse 190, CH-8057 Zurich, Switzerland}

\begin{abstract}
Identifying strongly connected substructures in large networks provides insight into their coarse-grained organization. Several approaches based on the optimization of a quality function, e.g., the modularity, have been proposed. We present here a multistep extension of the greedy algorithm (MSG) that allows the merging of more than one pair of communities at each iteration step. The essential idea is to prevent the premature condensation into few large communities. Upon convergence of the MSG a simple refinement procedure called ``vertex mover'' (VM) is used for reassigning vertices to neighboring communities to improve the final modularity value. With an appropriate choice of the step width, the combined \MVM algorithm is able to find solutions of higher modularity than those reported previously. The multistep extension does not alter the scaling of computational cost of the greedy algorithm.
\end{abstract}
\maketitle
\section{Introduction}
The networks under study in natural and social sciences often show a natural divisibility into smaller modules (or communities) originating from an inherent, coarse-grained structure. In general, these modules are characterized by an abundance of edges connecting the vertices within individual communities in comparison to the number of edges linking the modules. 

To detect these partitions several algorithm- or score-based approaches have been developed and applied. Very popular became the approach introduced by Girvan and Newman \cite{Newman2004} based on the quality function called ``modularity'' for partition assessment. This scoring function compares the actual fraction of intracommunity edges with its expectation in the random case given an identical degree distribution. The partition with the highest value of the scoring function is then considered to be the optimal splitting.  The modularity $Q$ is defined (for undirected networks) as
\[Q = \sum_{i=1}^{N_C} \left [ \frac{I(i)}{L} - \left ( \frac{d_i}{2 L} \right )^2 \right ] \]
with $I(i)$ the weights of all edges linking pairs of vertices in community $i$, $d_i$ the sum over all degrees of vertices in module  $i$, $L$ the total weight of all edges, and $N_C$ the number of communities. 

Intrinsically, the modularity based approach does not prescribe the usage of a particular optimization procedure. In practice, a strategy for optimization has to be chosen. The modularity optimization is a NP-hard problem \cite{Brandes2006}. Therefore, only an exhaustive search reveals the optimal solution for a generic network. This type of search is extremely demanding and only in a few cases feasible. Thus, many heuristic approaches such as extremal optimization \cite{Duch2005}, simulated annealing \cite{Guimera2005}, and the greedy algorithm \cite{Newman2004a} have been developed, refined, and successfully applied. Among the published approaches the greedy algorithm is one of the fastest techniques \cite{Danon2005}. On the other hand, many examples show that the greedy algorithm is not capable of finding the solutions with the highest modularity value. Furthermore, recent studies have provided evidence that modularity  \cite{Fortunato2007} and Potts model based approaches \cite{Kumpula2007} are endowed with an intrinsic resolution limit (small modules are not detected and amalgamated into bigger ones). Thus, each community has to be refined by subduing it as a separate network to the community detection algorithm. Therefore, a fast and accurate optimization technique is necessary.

In this article, we enhance the greedy algorithm by a multistep feature in combination with a local refinement procedure. The enhanced algorithm finds partitions with higher modularity values than previously reported.  This paper is organized as follows. In Sec.~\ref{algorithm} we introduce both procedures and describe the motivation for their construction. In addition, we discuss performance oriented implementations and estimate their running times.  Benchmarking results for a set of real-world networks and a comparison with other published results are presented in Sec.~\ref{benchmarks}. The conclusions are in Sec.~\ref{conclusions}. In this paper, all networks are considered as undirected. The extension to directed networks is straightforward. 

\begin{mybox}
 \hrule \vspace*{3pt}
 \begin{algorithmic}
\STATE Each vertex is a community
\STATE Calculate the modularity change matrix  $\Delta Q$ 
\STATE Determine the community degrees $d_i$
\WHILE{pair $(i,j)$ with $\Delta Q_{ij}> 0$ exists }
\FORALL{ element $(i,j,\Delta Q_{ij})$ in $\Delta Q$ matrix, parsed w.r.t. decreasing $\Delta Q$  and increasing $ (i,j)$}
\IF{ $\left \{ \begin{array}{l}
                \Delta Q_{ij} > 0 \mathrm{\;in \; best \;} l  \; \mathrm{values \;  in\; }\Delta Q \; \mathrm{matrix}\\
 i \mathrm{\; and \;} j\mathrm{\; unchanged\; in \; iteration} \\
 
               \end{array} \right \}$}
\STATE MergeCommunities(i,j)
\ENDIF
\ENDFOR
\ENDWHILE
 \end{algorithmic}
\hrule
\caption{Flowchart of the MSG algorithm. The modularity change is calculated according to Eq.~(\ref{merge-equation}). Details of the algorithm are given in algorithm \ref{details}. \label{concept}}
\end{mybox}

\section{The Algorithm}\label{algorithm}
\subsection{Multistep Greedy algorithm (MSG)}
\begin{figure*}
\includegraphics[width=14.5cm]{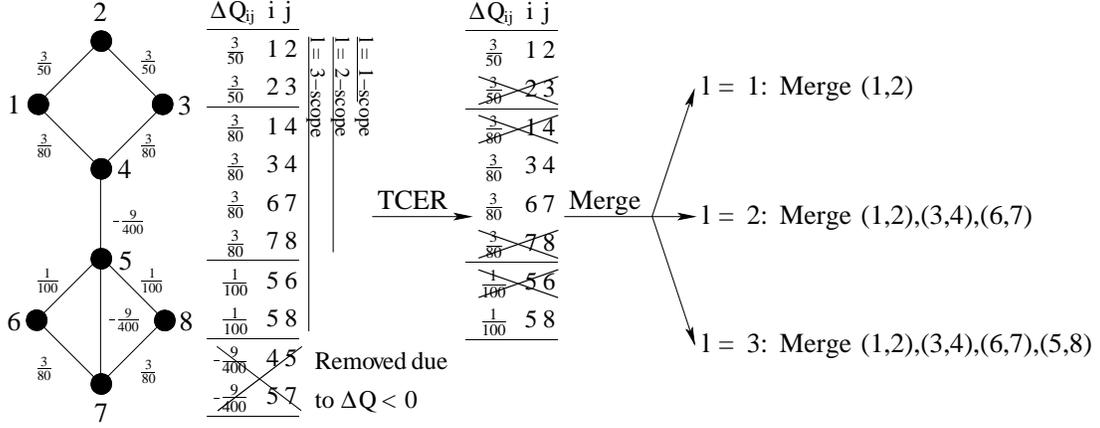}
\caption{Effect of different values of level parameter during first MSG-iteration on example network.\label{example}}
\end{figure*}

The classical greedy algorithm (first application in Ref.~\cite{Newman2004a}) joins iteratively the pair of communities that improves modularity most in each step. The essential idea of the ``multistep greedy'' algorithm (MSG) is to promote the simultaneous merging of several pairs of communities at each iteration. The pseudocode of the MSG algorithm is presented in algorithms \ref{concept} and \ref{details}, and an illustrative example is given in Fig.~\ref{example}. The MSG-algorithm starts with each vertex separated in its own community. At each iteration the modularity change $\Delta Q_{ij}$ upon merge of each pair of connected communities $(i,j)$ is calculated (while nonconnected pairs are ignored because their merging yields a negative modularity change). The triplets $(i,j,\Delta Q_{ij})$ are parsed in the order of decreasing $\Delta Q$-value and increasing community index. Those community pairs $(i,j)$ are joined which fulfill the following two criteria:
\begin{enumerate}
\item The modularity change $\Delta Q_{ij}$ is within the $l$ most favorable values (levels) and positive.
\item ``Touched-community-exclusion-rule'' (TCER): Neither module $i$ nor $j$ is present in another pair inducing a higher modularity change. 

\end{enumerate}
Convergence is reached when all pairwise merges of communities decrease modularity (by induction one can prove that all merges in further iterations would decrease modularity). A \textit{level} encompasses all triplets $(i,j,\Delta Q_{ij})$ with equal $\Delta Q_{ij}$-value and the \textit{level parameter} $l$ is kept constant. By construction the level parameter is always smaller than the number of edges in the network.

The multiple levels promote the concurrent formation of multiple centers. Simultaneously growing community centers hinder the condensation into few large communities (few formed communities scrape all vertices as the establishment of a new community is too expensive in modularity) as observed in the classical greedy algorithm. The TCER is a second mean against excessive aggregation into few large modules. This rule permits the addition of only one community to an existing community per algorithm iteration. Furthermore, the TCER guarantees that the modularity change upon all performed merges is just the sum over the corresponding $\Delta Q$ elements which improves efficiency.

\subsection{Implementation details of MSG} \label{implementation-MSG}
The key observation for an efficient implementation of the MSG is the following: Upon merge of communities $i$ and $j$ only those $\Delta Q$-elements concerning either of the two modules have to be recalculated. When the modules $i$ and $j$ are joined into a new one called $I$, the updated modularity changes $\Delta Q_{Ik}^{\rm new}$ (module $k$ is connected either to community $i$ or $j$) reads (see Sec.~II in Ref.~\cite{Clauset2004} for details) 
\begin{equation} \label{merge-equation}
 \Delta Q_{Ik}^{\rm new} = \left \{ 
\begin{array}{cl}
\Delta Q_{ik} + \Delta Q_{jk} & i,j \mathrm{ \; and \;} k\mathrm{\;pairwise \;connected} \\
\Delta Q_{ik} - \frac{d_j d_k}{2 L^2} & i \mathrm{\; and \; } k \; \mathrm{connected,}\; j \mathrm{\; and \; } k \; \mathrm{not} \\
\Delta Q_{jk} - \frac{d_i d_k}{2 L^2} & j \mathrm{\; and \; } k \; \mathrm{connected,}\; i \mathrm{\; and \; } k \; \mathrm{not} \\
\end{array} \right .
\end{equation}
with $d_x$ the sum over all degrees of vertices in community $x=i,j$ and $L$ the total edge weight.

Further efficiency improvements are gained from an appropriate choice of data structures. A \textit{set} (implementation taken from the \textsc{C++-STL}-library) is a sorted binary search tree. In a set individual elements can be found or inserted in $O(\log(n))$ time ($n$ the number of elements) and the extremal entries are found in constant time. The modularity changes are stored in the \textit{$\Delta Q$} matrix implemented as vector of row structures. The $i$th row consists of a set with elements $(j, \Delta Q_{ij})$ ($j$ a module linked to the community $i$) ordered according to the community index $j$. This data structure obsoletes a separate storage of the topology information. The extraction of the best $l$ modularity changes is handled via the \textit{level set}. For each pair of connected communities $i$ and $j$ the element $(\min\{i,j\},\max\{i,j\},\Delta Q_{ij})$ is added to the \textit{level set}. The \textit{level-set} elements are sorted with respect to decreasing $\Delta Q$ and increasing index values. The degree information is stored in a vector henceforth named \textit{d}. In each iteration a Boolean vector called \textit{touched} stores whether a community has already been modified in the same round. To save the time to determine the highest index of a present communities, the number of vertices (initial length) is chosen as length of the \textit{touched} vector. 

\begin{mybox}
\hrule \vspace*{3pt}
 \begin{algorithmic}
\STATE Each vertex is a community 
\STATE Calculate  community degrees $d$ and the $\Delta Q$ matrix 
\STATE Determine the initial modularity $Q\gets Q_0= - \sum_{i=1}^n \frac{d_i^2}{4 L^2} $
\STATE \textit{level set} $\gets $ set of $\Delta Q$ elements $(i,j,\Delta Q_{ij})$, sorted with respect to decreasing $\Delta Q$  and increasing $ (i,j)$

\WHILE{first element of \textit{level set} has $\Delta Q > 0$ }
\STATE $touched\gets (0,\ldots,0)$ Boolean, $N$-dimensional vector ($N = $ No.~vertices)
\STATE \COMMENT{$touched_i=1$, if module $i$ is modified in \textit{while}-loop}
\STATE $MP\gets$ subset of \textit{level-set} elements $(i,j,\Delta Q_{ij})$ with $\Delta Q_{ij} >0$ and $\Delta Q_{ij}$ among highest $l$ values
\FORALL{elements $(i,j,\Delta Q_{ij})$ of $MP$}

\IF{ (\textbf{not} $touched_i)$ \textbf{and} $(\mathbf{not\;} touched_j)$} 
\WHILE{parse $\Delta Q_{i.}$ and $\Delta Q_{j.}$ concurrently}
\STATE $\Delta Q_{ik}  \gets \left \{ 
\begin{array}{cl} 
\Delta Q_{ik} + \Delta Q_{jk} & i, k \; \mathrm{and}\; j,k \; \mathrm{are \; linked}\\
\Delta Q_{ik} - \frac{d_j d_k}{2 L^2} & i \mathrm{\; and \;} k \mathrm{\; are \; linked}  \\
\Delta Q_{jk} - \frac{d_i d_k}{2 L^2} & j \mathrm{\; and \; } k \mathrm{\; are \; linked}
\end{array} \right .  $
\STATE $\Delta Q_{ki}\gets \Delta Q_{ik}$
\STATE Update the \textit{level set} 
\STATE Update the modularity $Q\gets Q + \Delta Q_{ik}$
\ENDWHILE
\STATE Empty $\Delta Q_{j.}$
\STATE Flag $touched_i, touched_j\gets 1$
\STATE Update degrees: $d_i\gets d_i+d_j, d_j \gets 0 $
\ENDIF
\ENDFOR
\ENDWHILE
 \end{algorithmic}
\hrule
\caption{Performance-oriented implementation of MSG algorithm. The vector {\em touched} contains the information  for the touched-community-exclusion-rule (TCER). \label{details}}
\end{mybox}

The implementation details of the MSG algorithm are listed in algorithm \ref{details}. The calculation of the community degrees involves one parse of the edge information. In the second parse of the edge information the $\Delta Q$ matrix and the \textit{level set} is filled. The initial modularity change $\Delta Q_{ij}$ upon join of modules (at this stage the vertices) $i$ and $j$ is calculated as (see Sec.~II in Ref.~\cite{Clauset2004} for details)
\[ \Delta Q_{ij} = \frac{I}{L} - \frac{d_i d_j}{2 L^2}\]
with $I$ the weight of the edges connecting the vertices $i$ and $j$, $d_{x}$ the degree of vertex $x=i,j$, and $L$ the total edge weight. The modularity value of the initial partition is ($N$ the number of vertices)
\[ Q_0 = - \sum_{i=1}^N \frac{d_i^2}{4 L^2}.\]
The algorithm iteration starts by initializing the \textit{touched} vector. Subsequently, the \textit{Level}-set is parsed and all elements with positive $\Delta Q$ value, whose modularity change is among the best $l$ (external \textit{level parameter}) different values, are stored in a set named \textit{MP} conserving the order of the \textit{level set}. In this order the module pairs are merged unless one of them was part of a amalgamation in the same algorithm iteration. In the merge process, the changed $\Delta Q$ matrix elements are calculated as described at the beginning of this paragraph. To determine which case applies in Eq.~(\ref{merge-equation}) the fact that each row of the $\Delta Q$ matrix is ordered with respect to the community index can be used. More precisely, parse for the merge of modules $i$ and $j$ the corresponding rows concurrently. For each row define an momentarily considered element $p$. If the community index of $p_i$ is equal to the one of $p_j$, the first case applies and advance both $p$'s to the next element in the corresponding row. If the index $k$ of $p_i$ is lower than the one of $p_j$ calculate the $\Delta Q_{Ik}^{\rm new}$ element ($I$ the name of the merged community) according to the second case and advance (if possible) only $p_i$. If the module index of $p_i$ is larger than the one of $p_j$, proceed analogously. If one $p$ reaches the end of the row, merge the remaining elements of the other row according to the respective rule. This procedure will be called ``asynchronous parsing'' in Sec.~\ref{complexity-MSG}. It is customary to update each $\Delta Q$ element after calculation. To complete the merge process it remains to update the community degrees and to flag the modified communities in the \textit{touched} vector.

\subsection{Running time estimation of MSG} \label{complexity-MSG}
As we adopted the modularity change calculation of Clauset et al. (Sec.~II in Ref.~\cite{Clauset2004}) we can adopt their method of running time estimation as well. First, we observe that the update of one element in the $\Delta Q$ matrix and the \textit{level set} costs in the worst case $O(\log(N))$ (insertion in set, each community has at most $N$ neighbors with $N$ the number of vertices) and $O(\log(M))=O(\log(N))$ running time (the number of distinct edges $M$ is bounded by the square of the number of vertices $N^2$), respectively. 

Merging communities $i$ and $j$ involves an update of the $\Delta Q$ matrix and the \textit{level set} for each element of the corresponding rows of the $\Delta Q$ matrix . The calculation of each changed value can be achieved in constant time as during the asynchronous parsing it is known whether the other community is linked as well and all other information (community degrees) is stored in a vector. Thus, the total running time contribution of one merging event is $O((d_i + d_j) \log(N))$ with $d_k$ the number of edge starts/ends on vertices of community $k = i,j$. In the worst case all communities are changed in one algorithm round. As the sum over all $d_i$ values is twice the number of distinct edges, the contribution of the merging processes in one algorithm round is at most $O(M \log(N))$. 
The other steps of one algorithm round are less consumptive: The extraction of pairs belonging to the best $l$ levels can be performed in constant time. The same is true for the update of the degree information. 
If $D$ is defined as the depth of the dendrogram of communities, at most $D$ algorithm rounds have to be performed. Thus, the running time expectation for the iterative part is $O(D M \log(N))$ which is identical to the complexity of the classical greedy algorithm \cite{Clauset2004}. 

The initialization involves the read-in processes of the edge information ($M$ constant time operations), the degree calculation (part of read-in process), the calculation of the initial modularity (constant time operation on $N$ elements) and finally the generation of the $\Delta Q$ matrix and the \textit{level set} at costs $O(M \log(N))$ ($M$ insertions in a set with at most $N$ or $M$ elements, respectively). In the worst case the expected contribution of the initialization to the running time is $O(M \log(N))$. 

In the precedent paragraphs we have shown that the MSG greedy algorithm has the total complexity $O(D M \log(N))$. Among the published strategies for modularity optimization the classical greedy algorithm \cite{Clauset2004} is the fastest \cite{Danon2005}. As the MSG shares the worst case expectation for the running time with the classical greedy algorithm, we conclude that the MSG is one of the fastest procedures for modularity optimization. 

\subsection{Vertex mover (VM)}
To further improve modularity by ``adjusting'' misplaced vertices, a refinement step called ``vertex mover'' (VM) is applied upon convergence of the MSG algorithm. In principle, it could also be applied to other modularity optimization procedures. In the VM, the list of vertices is parsed in the order of increasing degree and vertex index (to resolve the degeneracy of multiple vertices with equal degree) and every vertex is reassigned to the neighboring community with maximal modularity improvement. This parsing-and-reassignment procedure is repeated until no modularity improvement is observed. 

The VM procedure is similar to the Kernighan-Lin algorithm \cite{KernighanLin72} (applied to modularity optimization in Ref.~\cite{Newman2006a}). In contrast to the Kernighan-Lin algorithm the VM procedure has a perfectly local focus. In other words, instead of repetitively searching for the optimal vertex to reassign, the VM procedure parses the vertices in the aforementioned order and identifies the optimal community for the considered vertex. Furthermore, each reassignment of the VM approach improves modularity. Therefore, the selection of the optimal intermediate partition as in the Kernighan-Lin algorithm is not necessary.

\begin{figure*}
\begin{tabular}{c}
\includegraphics[width=13.5cm]{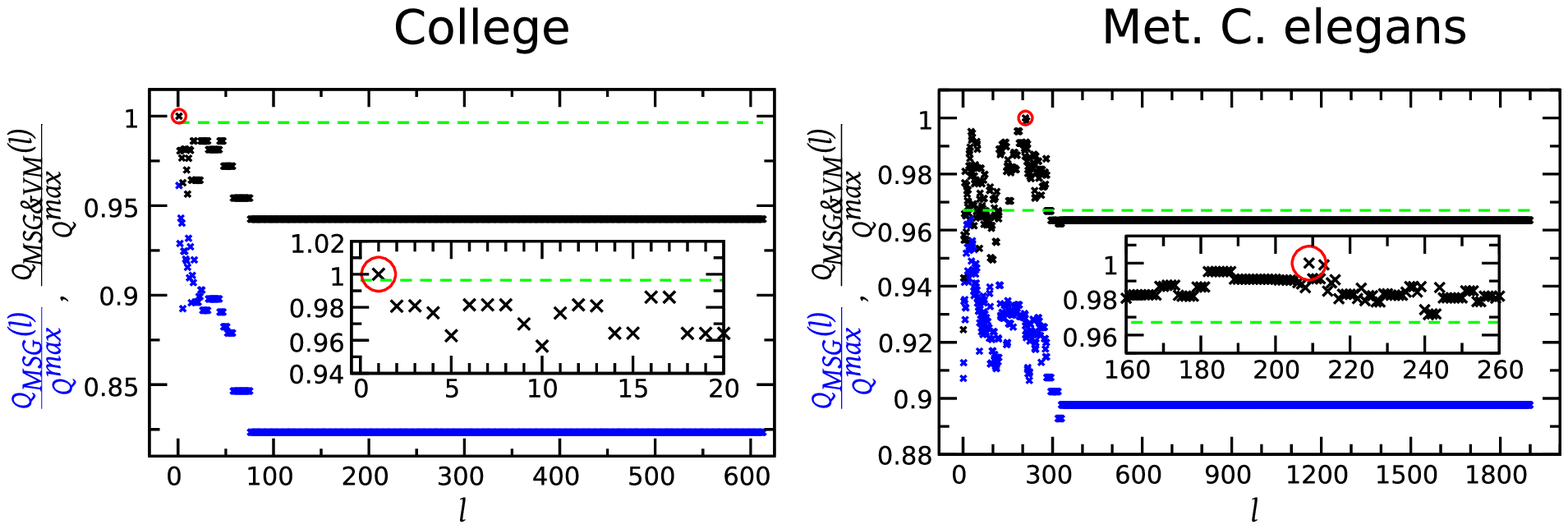}\\[1.6cm]
\includegraphics[width=13.5cm]{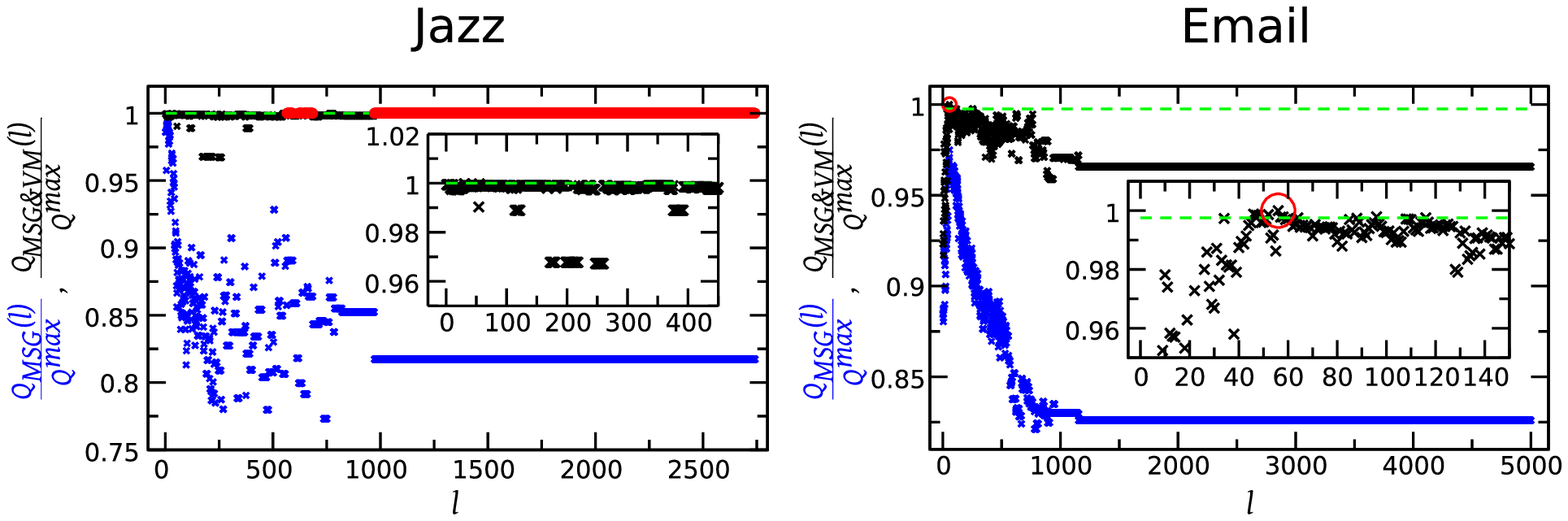}\\[1.6cm]
\includegraphics[width=13.5cm]{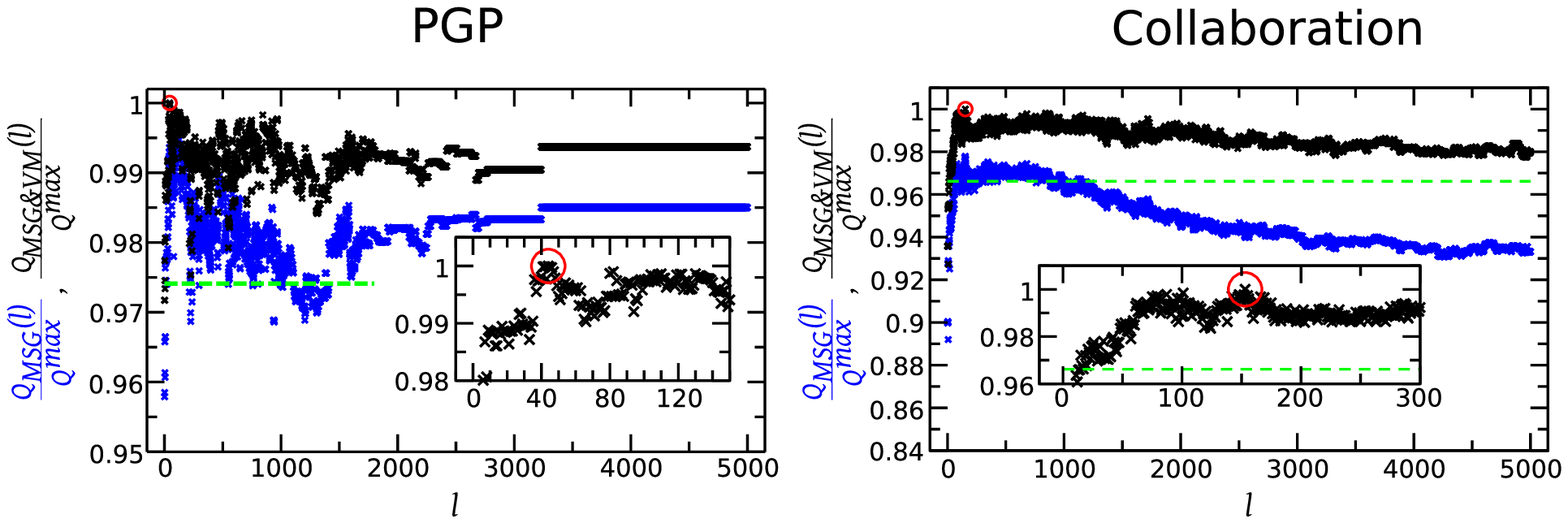} \\[1.0cm]
\end{tabular}

\caption{\label{profiles}(Color online) Dependence of MSG modularity value $Q_{\rm MSG}(l)$ (blue), \MVM modularity value $Q_{\mbox{MSG-VM}}(l)$ (black) on the level parameter $l$ relative to maximal \MVM modularity value $Q^{\rm max}$. The previously published result $Q_{\rm pub}/Q^{\rm max}$ (dashed green line) is also shown as basis of comparison. The red circles indicate the value of $l$ that yields maximal modularity. A significant number of $l$-values yield higher modularity than the previously published maximal modularity for all but the smallest two networks, i.e. Zachary (not shown) and College. In the latter, only $l=1$ yields a higher modularity than $Q_{\rm pub}$.  }
\end{figure*}

\subsection{VM implementation} \label{performance-VM}
The modularity change $\Delta Q$ upon reassignment of vertex $v$ from community $i$ to $j$ can be written as 
\begin{equation} \label{formula}
\Delta Q = \frac{\mathrm{links}(v \leftrightarrow j) - \mathrm{links}(v \leftrightarrow i)}{L} - \frac{k_v \left ( d_j - d_{i \setminus v}\right ) }{2 L^2}
 \end{equation}
with $k_v$ the degree of vertex $v$, $d_j$ the sum over the degrees of all vertices in community $j$, $d_{i \setminus v} = d_i - k_v$ the corresponding degree for community $i$ without vertex $v$, and $L$ the total weight of all edges.

The most time consuming part of the VM is the calculation of the modularity changes upon reassignment of the vertices. Consequently, Eq.~(\ref{formula}) reduces this bottleneck to the calculation of weight of the edges connecting the vertex to the neighboring communities. {\color{black} The connectivity information of vertex $v$ is stored in a sparse vector [i.e., a vector of elements $(u,w_{vu})$ with $u$ a vertex linked to $v$ and $w_{vu}$ the total weight of all edges connecting vertices $u$ and $v$]. These rows are stored in a vector and form the topology matrix. To determine the total edge weight connecting vertex $v$ with community $j$ the $v$th row is parsed and for each entry the weight is added to the subtotal edge weight of the corresponding community. To keep access times short a $N$-dimensional vector ($N$ the number of vertices) is chosen to store the intermediate $\mathrm{links}(v \leftrightarrow j)$ results. The optimal reassignment partner for vertex $v$ is the community with smallest index yielding the maximal modularity improvement.
}

\subsection{Estimation of VM running time} \label{complexity-VM}
Calculating the modularity changes upon reassignment of one vertex to any neighboring community involves one parse of its edge list supplemented with direct memory access to determine the community affiliation and some constant time operations for the actual modularity calculation. Therefore, the running time contribution of one vertex is proportional to its degree. One algorithm round requires $O(L) = O(\sum_i d_i)$ running time. The estimation of the number of needed iterations is not possible as it depends on the quality of the MSG result. In all examples tested by us the running time of the VM was always at least one order of magnitude smaller and less than one minute even for the biggest networks under study.

\begin{table*}
\hspace*{-10mm}
\begin{tabular}{lcrrrrrrrrrrr}\hline \hline 
\multicolumn{4}{c}{Network} &\hspace*{7mm} &   \multicolumn{4}{c}{MSG-VM} & \hspace*{7mm} & \multicolumn{3}{c}{Greedy} \\ \cline{1-4} \cline{6 -9} \cline{11 -13}
Name &  Ref. & Vertices & Edges  & &  $l_{\rm opt}$ & \multicolumn{1}{c}{\hspace*{4mm}$Q$\hspace*{4mm}} & Time $[s]$ &$N_C$ & & \multicolumn{1}{c}{$Q$} & Time $[s]$ & $N_C$ \\ \hline
Zachary Karate Club   &   \cite{Zachary1977}  &  34   &   78 &    &  3   & 0.398  &   na & 4 & &   0.381  &   na  & 3 \\
Metabolic \textit{E.~coli}   &  \cite{Ma2003}  &   443   &   586  &   &    6,    8  &   0.816  &   na  & 19 &  & 0.811   &   na  & 20\\
College   Football    &  \cite{Girvan2002}    &    115    &   613    &   &   1  &   0.603  &   na  & 8 & &  0.556  &  na & 6   \\
Metabolic  \textit{C.~elegans}  &  \cite{Jeong2000}    &   453   &   1899    &    &    209  &   0.450  &   na   & 8 &  &   0.412   &  na  & 13 \\
Jazz   &   \cite{Gleiser2003}  &  198  &   2742   &   &   566   &  0.445  &   na  & 4 & &  0.439  &   na & 4  \\
Email  &   \cite{Guimera2003}  &  1133   &   5451  &   &  56  & 0.575  &   na  & 10 & & 0.503  &   na  & 12 \\
Yeast  (PPI,  CP)  &  \cite{Krogan2006}   &   2552  &   7031  &     &  35    &  0.706  &   na  & 33 &  & 0.675   &  na  & 51 \\
M.~Karplus   &   \cite{karplus}    &  1167   &   13423  &   &  91  &   0.316  &   na  & 11 &  & 0.264  &   na   & 18 \\
PPI-CP    \textit{S.~cerevisiae}   &  \cite{Colizza2005}  &   4626  &   14801 &     &    170  &   0.545  &   na  & 24 &  & 0.500   &  na  & 38 \\
PPI  \textit{S.~cerevisiae}   &  \cite{Colizza2005}  &   4713  &   14846  &  &    170   &   0.546  &   na  & 65 & &  0.501   &  na  & 81\\
M.~Karplus weighted    &  \cite{karplus}  &    1167   &   18991 &       &   173 &   0.320  &   na  & 13 & & 0.296  &  na  & 11 \\
Internet  &   \cite{AS2001}  &  11174  &   23409    &   &  278    &  0.625  &   8   & 35 & &  0.584  &   8 & 49\\
PGP-key signing  &   \cite{Guardiola2002,Boguna2004}  &  10680  &   24340  &     &   44 &  0.878  &   2   & 140 &  &  0.849  &   3 & 195\\
Word Association  (CP)   &  \cite{Nelson2004}    &    7204   &   31783 &     &   71  &   0.541  &   4   & 16 & &  0.452  &  7 & 52\\
Word  Association &   \cite{Nelson2004}   &  7207   &   31784    &   &   97  & 0.540  &   3   & 17 & &  0.465  &   7 & 38\\
Collaboration  &   \cite{Newman2001b}  &  27519  &   116181 &    &   153   &  0.748  &   14  &82 &   & 0.661  &   103 & 381\\
WWW  &   \cite{Albert1999}   &  325729    &   1117563 &    &   3034  &  0.939  &   562  & 674  &  & 0.927  &   7640 & 2183\\
Actor  &   \cite{Barabasi1999}    &  82583  &   3666738 &    &   2429  &  0.543  &   1722   & 238 &  &  0.470  &   6288 & 406\\
Actor  weighted    &  \cite{Barabasi1999}  &    82583  &   4475520 &     &   389 &   0.536  & 5099 & 322  &  &   0.480  &  3541 & 361\\ \hline \hline 
\end{tabular}
\caption{\label{real-world}Results on real-world examples. Among all tested level parameters \rc{(all positive integers smaller than 5000 or the number of edges if smaller)} the value $l_{\rm opt}$ yields the highest value of  $Q$ for the considered network. $N_C$ is the number of communities found. In most cases, a larger number of communities (larger $N_C$) is identified by the classical greedy than the \MVM extension because the former partitions the network in few large communities and many small communities with less than ten vertices (mostly 2 - 20 times more small modules identified by greedy than MSG-VM). The \MVM approach prevents the condensation into few large modules: The three largest modules contain between 1.5 and 4 times less vertices in the \MVM partition than in the greedy partition (not shown).  The running time (on a recent laptop) is reported for a single run of the algorithm. The entry ``na'' indicates that the running time is shorter than 1 s and therefore not displayed. The suffix ``CP'' points out that only the largest connected component (the ``central part'') was considered. The acronym ``PPI'' stands for ``protein-protein interaction.''}
\end{table*}

\begin{table}
\begin{tabular}{lllcc} \hline \hline 
Network  &   $Q_{\rm max}^{\mbox{\tiny MSG-VM}}$  &   $Q_{\rm pub}$  &  Source & Method\\ \hline
Zachary Karate Club  &   0.398  &   0.419  &   \cite{Newman2006a}  &   \cite{Newman2006a}\\
College   Football    &    0.603  &   0.601  &   \cite{Girvan2002}  &  \cite{Girvan2002}\\
Metabolic    \textit{C.~elegans}  &   0.450  &   0.435    &   \cite{Newman2006a}    &  \cite{Newman2006a}\\
Jazz    &   0.445  &   0.445  &   \cite{Newman2006a}  &   \cite{Duch2005}\\
Email   &   0.575  &   0.574  &   \cite{Newman2006a}  &   \cite{Duch2005}\\
PGP-key signing  &   0.878  &   0.855  &   \cite{Newman2006a}  &   \cite{Newman2006a}\\
Collaboration  &   0.748  &   0.723  &   \cite{Newman2006a}  &   \cite{Newman2006a}\\ \hline \hline
\end{tabular}
\caption{\label{performance-comparision}Comparison of maximal value of modularity obtained by the \MVM algorithm $Q_{\rm max}^\mathrm{\MVM}$ with previously published results $Q_{\rm pub}$. The highest published value was extracted from the referenced paper (``Source'') where it has been calculated by the ``Method'' whose reference is listed in the last column.}
\end{table}

\section{Results} \label{benchmarks}
\subsection{Test set of networks} \label{benchmarking-sets}
For benchmarking algorithms that optimize modularity the networks commonly used are the collaboration network (coauthorships in cond-mat articles) \cite{Newman2001b}, the graph of metabolic reactions in \textit{Caenorhabitis elegans} \cite{Jeong2000}, the email network \cite{Guimera2003}, the network of mutual trust (PGP-key signing) \cite{Guardiola2002,Boguna2004}, the conference graph of college football teams \cite{Girvan2002}, the network of jazz groups with common musicians \cite{Gleiser2003} and the Zachary karate club example \cite{Zachary1977}.  In addition, we include less frequently used examples such as the graph of the metabolic reactions in \textit{Escherichia coli} \cite{Ma2003},  two different data set describing the protein-protein interactions in S. cerevisiae (budding yeast) \cite{Krogan2006,Colizza2005} with labels ``PPI'' and ``yeast.'' To cover linguistic applications we benchmark the word association network \cite{Nelson2004} and the graph of the co-appearing words in publication titles (co)authored by Martin Karplus \cite{karplus} who has the third highest $h$-factor \cite{Hirsch2005} among chemists  \cite{Ball2007}. Further aspects of social webs were incorporated by considering the graph of costarring actors in the IMDB database \cite{Barabasi1999}. Noticeable, the actor network - being the network with the largest number of edges - serves as a proof of concept for such big networks being treatable as well.  From computer science we include the internet routing network \cite{AS2001} and the graph of World Wide Web pages \cite{Albert1999}. With this selection of networks most currently known application fields of networks are covered. To study the effect of disconnected graphs and weighted networks, we consider in both cases the full network as well as the largest connected component (suffix ``CP'') and the unweighted variant, respectively. Unless stated otherwise the networks are treated unweighted.

\subsection{Dependence on $l$ and vertex labeling }

It is important to investigate the robustness upon the choice of $l$ and to determine the highest modularity values achievable with the \MVM algorithm. There is a minor dependence on the value of $l$ (Fig.~\ref{profiles}) which changes the \MVM modularity by less than 2 \% for large networks. Moreover, the maximal modularity is obtained with $l < 300 $ for 14 of the 19 networks (Table \ref{real-world}). An empirical formula for the optimal choice of the level parameter will be presented elsewhere. 

Noteworthily, for a labeled graph and a chosen level parameter the algorithm is deterministic. To assess the contribution of the labeling, the benchmarking procedure is performed also on hundred copies of the smallest ten networks with permuted vertex labels. This permutation leaves the topology invariant, but modifies the order in which the community pairs are considered. In comparison to the maximal modularity value found for the unscrambled variants a maximal improvement of 0.94 \% is observed. 

\subsection{Performance and running time} \label{performance-results}
The modularity values obtained with the \MVM approach are listed in Table \ref{performance-comparision}. For five of the seven networks considered here the \MVM algorithm finds solutions with modularity higher than previously published. Only for the Zachary Karate network the \MVM procedure  yields a smaller modularity value. For the jazz network a solution with the identical $Q$ value is obtained. For the networks without published modularity values we compare the optimal values obtained by the \MVM algorithm with the classical greedy algorithm for modularity optimization as introduced by Newman \cite{Newman2004a}  in Table \ref{real-world}. We observe that the \MVM algorithm outperforms the original greedy algorithm significantly. 

The running time estimations in Secs.~\ref{complexity-MSG} and \ref{complexity-VM} are based on a worst case scenario. To investigate the running time behavior on real-world examples, we compare the running times of the classical greedy variant and the \MVM algorithm in Table \ref{real-world}. These data show that given the appropriate level parameter choice the \MVM algorithm is in almost all cases faster than the classical greedy algorithm and, at the same time, reaches a higher value of modularity. 

\section{Conclusions} \label{conclusions}
To prevent premature condensation into few large communities the greedy algorithm for modularity optimization has been extended by a procedure for simultaneous merging of more than one pair of communities at each step. Furthermore, this ``multistep'' greedy variant has been combined with a simple vertex-by-vertex a posteriori refinement. On seven networks with previously published modularity values the \MVM algorithm combination outperforms all other frequently used, generic techniques except for the smallest of the seven examples. In addition, a single run of the \MVM algorithm requires similar computer time as the greedy algorithm. In most cases less than 10 independent (i.e., embarrassingly parallel) runs of \MVM are required to obtain a modularity within 1 \% of the highest value because an empirical formula has been derived for the appropriate choice of the optimal step-width. Therefore, the \MVM algorithm is an efficient tool to find network partitions with high modularity \footnote{The code is available at \textit{http://www.biochem-caflisch.uzh.ch/communitydetection/}}.

\section{Acknowledgments}
The authors thank Stefanie Muff and Francesco Rao for helpful discussions. Christian Bolliger, Thorsten Steenbock, and Dr.~Alexander Godknecht are acknowledged for maintaining the Matterhorn cluster where most of the parameter studies were performed. We are thankful to Drs.~Arenas, Barab\'{a}si, Gleiser, and Newman for providing the network data. This work was supported by a Swiss National Science Foundation grant to A.C.



\end{document}